\documentclass[12pt]{article}
\usepackage{indentfirst}
\usepackage{graphicx}
\usepackage{epsfig}
\usepackage{latexsym}
\usepackage{amsmath}
\usepackage{amssymb}
\usepackage{amsfonts}
\usepackage{mathrsfs}
\usepackage{pifont}
\usepackage{amsthm}
\usepackage{listings}
\usepackage[colorlinks=true,linkcolor=blue]{hyperref}
\usepackage{tabularx}
\usepackage{ulem}
\usepackage[tmargin=2cm,bmargin=2cm,lmargin=2cm,rmargin=2cm]{geometry}
\usepackage{float}
\usepackage{cases}
\usepackage{subcaption}
\usepackage{mathtools}
\usepackage{underscore}
\usepackage{setspace}
\usepackage{multicol}
\usepackage{xcolor}
\usepackage{cite}
\usepackage{bigstrut}
\usepackage{placeins}
\usepackage[utf8]{inputenc}
\usepackage{framed}

\usepackage{autobreak}
\newcommand{\dd}{\mathrm{d}}
\newcommand{\ii}{\mathrm{i}}

\newcommand{\mbf}{\boldsymbol}
\newcommand{\pl}{\partial}
\newcommand{\pdif}[2]{\dfrac{\partial #1}{\partial #2}}

\pdfoutput=1

\title{Cell motility modes are selected by the interplay of mechanosensitive adhesion and membrane tension}
\author{Yuzhu Chen, David Saintillan, Padmini Rangamani}
\date{\today} 
\begin{document}
\maketitle

\begin{abstract}
    The initiation of directional cell motion requires symmetry breaking that can happen both with or without external stimuli. 
    During cell crawling, forces generated by the cytoskeleton and their transmission through mechanosensitive adhesions to the extracellular substrate play a crucial role. 
    In a recently proposed 1D model (Sens, PNAS 2020), a mechanical feedback loop between force-sensitive adhesions and cell tension was shown to be sufficient to explain spontaneous symmetry breaking and multiple motility patterns through stick-slip dynamics, without the need to account for signaling networks or active polar gels. 
    We extended this model to 2D to study the interplay between cell shape and mechanics during crawling. 
    Through a local force balance along a deformable boundary, we show that the membrane tension coupled with shape change can regulate the spatiotemporal evolution of the stochastic binding of mechanosensitive adhesions. 
    Linear stability analysis identified the unstable parameter regimes where spontaneous symmetry breaking can take place. 
    Using simulations to solve the fully coupled nonlinear system of equations, we show that starting from a randomly perturbed circular shape, this instability can lead to keratocyte-like shapes.\ 
    Simulations predict that different adhesion kinetics and membrane tension can result in different cell motility modes including gliding, zigzag, rotating, and sometimes chaotic movements. 
    Thus, using a minimal model of cell motility, we identify that the interplay between adhesions and tension can select emergent motility modes.
\end{abstract}

The mechanism of cell crawling on substrates is important for understanding numerous biological processes such as morphogenesis and wound healing \cite{holt2021spatiotemporal}.
Experiments have revealed that several main subcellular processes, including actin polymerization \cite{pollard2003cellular}, adhesion \cite{ananthakrishnan2007forces}, and myosin contraction \cite{yam2007actin}, are spatially and temporally orchestrated to generate coherent cellular motion. 
These experiments have also lent themselves to systematic theoretical and computational modeling \cite{mogilner2020experiment}. 
Among these different subprocesses, the initiation of motion is of particular interest because it can happen both due to external cues and spontaneously due to intrinsic biochemical or mechanical instabilities, in a process known as cell self-polarization \cite{cramer2010forming,rappel2017mechanisms}. 
A polarized cell undergoes distinct molecular processes at the front and rear, such as the distribution of Rho family GTPases which regulate the actin protrusion and adhesion formation \cite{ridley2003cell}; this distribution specifies a direction for motility. 
Interestingly, despite the vast number of molecular players involved \cite{ridley2003cell,parsons2010cell}, cell migration is essentially a mechanical process \cite{sens2020stick,rangamani2011}. 
The integration of cell signaling into mechanical processes has led to different scales of biophysical models \cite{danuser2013mathematical,rangamani2011,Xiong2010}. 

Depending on the cell type, migrating cells can assume different shapes. 
For example, fibroblasts have multiple protrusions \cite{lo2000cell,Dubin-Thaler2004,Dubin-Thaler2008}, while fast-moving keratocytes are characterized by their flat, smooth, fan-shaped leading edges \cite{keren2008mechanism}. 
Despite these differences, a common mechanism for cells to undergo directional motion can be summarized as follows:
the leading edge of the cell protrudes as a result of the speed difference between actin polymerization and retrograde flow, the rear retracts to keep up with the front \cite{yam2007actin,wilson2010myosin}, and membrane tension plays a crucial role in this process because it can coordinate the protrusion and retraction as a global regulator for cell shape change and motility \cite{lieber2013membrane,diz2013use,sens2015membrane}. 
With this picture, many phenomenological models have been proposed to explain the underlying mechanisms for motility and shape determination. 
These include models constructed based on the graded radial extension hypothesis \cite{rubinstein2005multiscale}, viscoelastic actin network and myosin transport \cite{rubinstein2009actin}, force balance between treadmilling actin filaments and membrane tension \cite{schaus2007self,keren2008mechanism}, two-phase fluids with actin polymerization \cite{herant2010form}, and many other redundant mechanisms \cite{wolgemuth2011redundant}. 
Later, more comprehensive free-boundary models incorporated other features such as the discrete stick-slip adhesions \cite{shao2010computational,shao2012coupling}, the orientational order of the actin filament network \cite{ziebert2012model,tjhung2012spontaneous,tjhung2015minimal,marth2015mechanism}, and the feedback loop between actin flow, myosin, and adhesion \cite{barnhart2015balance} and are  reviewed in \cite{holmes2012comparison}. 
Most of these simulations either start with a crescent shape to match experimental observations \cite{rubinstein2005multiscale, rubinstein2009actin}, with perturbations or polarity fields along a specified direction \cite{shao2010computational,shao2012coupling,ziebert2012model,tjhung2012spontaneous,tjhung2015minimal,marth2015mechanism,barnhart2015balance}, or with a prescribed front \cite{herant2010form}.
Therefore, how the cell shape transitions from random fluctuations to persistent motile shapes remains unclear from these models. 

\begin{figure*}[t]
            \centering
            \includegraphics[width=0.92\textwidth]{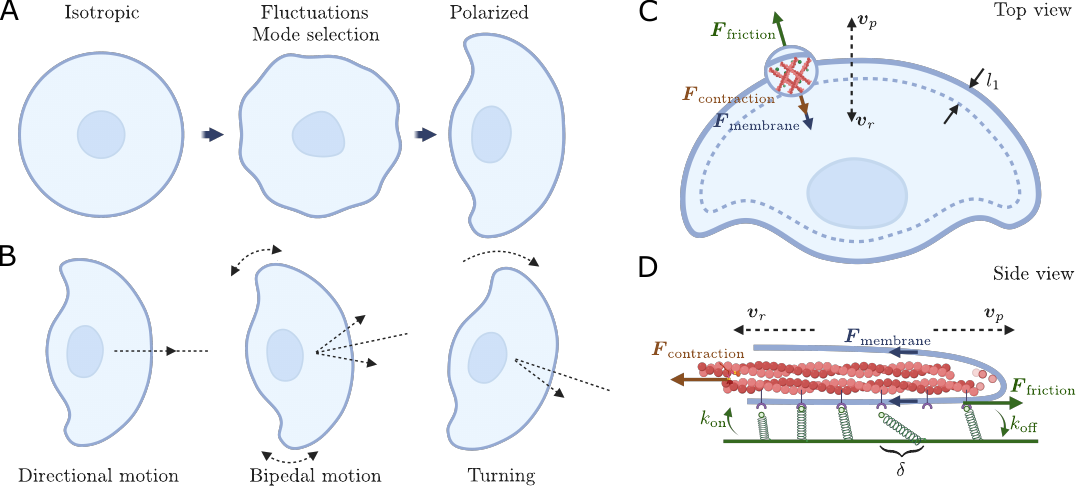}\vspace{-0.1cm}
            \caption{(\textit{A}) Schematic describing self-polarization in cells. 
            Fluctuations of certain wavelengths are amplified by the intrinsic instability of the feedback loop between adhesions and membrane tension, leading to self-polarization. 
            (\textit{B}) Schematic describing the various motility modes. 
            Directional motion is the uniaxial motion of the cell in one preferred direction. 
            Bipedal motion involves antiphase retraction of the left-right trailing edge and lateral oscillation of the cell.  
            In turning motion, the cell rotates in a certain direction with left-right asymmetry. 
            (\textit{C}) Sketch of the two-dimensional model (top view). The cell shape is determined by the difference between the polymerization velocity $v_p$ and the retrograde velocity $v_r$, where $v_r$ is related to the off-rate of adhesive bonds distributed within the lamellipodium with width $l_1$ along the cell boundary. 
            The force balance between friction, membrane tension, and contraction is maintained at the cell edge. 
            (\textit{D}) Sketch of the two-dimensional model (side view). 
            The adhesive bonds bind to the substrate with a constant rate $k_{\text{on}}$ and unbind with a force-dependent rate $k_{\text{off}}$ as the actin filaments move.}
            \label{fig:schematic}\vspace{-0.2cm}
\end{figure*}

Apart from steady-moving states, keratocytes can also undergo more complex motility modes such as bipedal motion and spontaneous turning, or a combination of both \cite{oliver1999separation,ream2003influences,barnhart2010bipedal,loosley2012stick}. 
Experimental and modeling studies reveal that the stick-slip adhesive sites at the rear and their coupling with the cytoskeleton dynamics are crucial for such unsteady motions \cite{barnhart2010bipedal,loosley2012stick,allen2020cell,lee2020modeling}. 
However, the models used to investigate this coupling need prior knowledge of the positions of adhesion sites and the broken symmetries.
As a result, they cannot predict how the distributions of adhesions are formed in the first place. 
Other simulations that obtain bipedal motion without prescribed adhesion sites take the substrate deformations into account \cite{ziebert2013effects,lober2014modeling}, but treat adhesion dynamics phenomenologically.

Mogilner \textit{et al.} suggested that cell polarization and turning may share similar mechanisms that involve the feedback between actomyosin flows and stick-slip dynamics of adhesions \cite{mogilner2020experiment}. 
Recently, Sens proposed a 1D mechanical feedback loop between the binding and unbinding of cell-substrate adhesions and linear cell tension that can lead to the spontaneous symmetry breaking \cite{sens2020stick}. 
Based on a mean-field approximation of the molecular clutch model for adhesions, this theoretical model also predicted various one-dimensional cell locomotion behaviors such as steady crawling, bistability, and bipedal motion as observed in experiments \cite{barnhart2010bipedal,loosley2012stick}, bridging the gap between microscopic factors and whole cell locomotion.\ 
Combining the mechanosensitive unbinding of adhesions \cite{sens2020stick} that are based on first principles, with cell shape change can provide a natural description for the formation and the stick-slip dynamics of adhesions and how they can lead to unsteady motions. 
However, this model does not explore the link between the mechanical feedback loop and cell shape and motility modes. 
Such an exploration requires a 2D formulation with a deformable boundary.
Here, we asked whether the mechanical feedback loop between adhesion and cell shape change is sufficient to explain the dynamics of cell shape change and motility modes. 
Specifically, we extend the 1D model in \cite{sens2020stick} to two dimensions with deformable boundaries to investigate how the coupling between membrane tension and the stick-slip dynamics of adhesions determines cell shape change and the spatial distribution of traction forces during the initiation of motility and resulting sustained motion. 
Linear stability analysis of the model shows that uniform steady states are unstable in certain parameter regimes due to the stick-slip nature of the adhesions.
Numerical simulations in 2D predict that the interplay between membrane tension and mechanosensitive adhesions is sufficient for a circular-shaped cell to spontaneously initiate and sustain motion  (Fig.~\ref{fig:schematic}\textit{A}), and the various motility modes mentioned above can be captured (Fig.~\ref{fig:schematic}\textit{B}) even without the consideration of the complex reorganization of the cytoskeleton. 
Thus, our 2D model is able to capture the initiation of cell polarization and the different motility modes (directional, bipedal, and turning) by tuning the physical parameters, and demonstrates the crucial role for interplay between adhesions and membrane tension in cell motility.

\section*{Model Development}
\subsection*{Governing equations}
        At cellular length scales, inertia is negligible \cite{danuser2013mathematical}, so that forces are balanced everywhere at any instant of time. 
        When the actin filaments treadmill at the edge of the cell, the membrane imposes an opposing force while actomyosin contraction generates a contractile force (Fig.~\ref{fig:schematic}\textit{C}). 
        These two forces lead to the retrograde flow of actin filaments away from the cell edge and are locally balanced by a friction force pointing outwards and created by transmembrane adhesions that can stochastically bind or unbind to the flat substrate \cite{schwarz2013physics}: 
            \begin{align}
                \mbf{F}_{\text{contraction}} + \mbf{F}_{\text{membrane}} + \mbf{F}_{\text{friction}} = \mbf{0}.
            \end{align}
        Here, we apply a coarse-grained approach and assume the focal adhesions are concentrated in a narrow region near the cell periphery with width $l_1$, as proposed by \cite{sens2020stick}. 
        The cell boundary is thus treated as a one-dimensional curve $\Gamma(t)$ evolving in a two-dimensional plane (see Fig.~S1 of the \textit{SI Appendix}), where the adhesion clusters are material points containing a collection of $\rho l_1$ adhesive linkers on average per unit length along the curve. 
        The position vector of a material point on the boundary at time $t$ is represented by $\mbf{x}(\alpha,t)$, where $\alpha\in[0,2\pi]$ is a Lagrangian parameter (a detailed description of the geometry and parameterization can be found in the \textit{SI}). 
        At each material point, the total number of adhesive linkers for each adhesion cluster is assumed to be conserved \cite{schwarz2013physics}. 
        Each of the adhesive linkers can unbind and rebind between the substrate and the sliding actin filaments with an off-rate $k_{\text{off}}$ and an on-rate $k_{\text{on}}$ (Fig.~\ref{fig:schematic}\textit{D}), and the effect of thermal fluctuations is modeled with an effective diffusivity $D$. 
        Denoting the current arclength as $s(\alpha,t)\in[0,L(t)]$, the fraction of bound linkers $n(\alpha,t)$ at each material point evolves by the kinetic equation
        \begin{align}
            \pdif{n}{t}(\alpha,t) = k_{\text{on}}(1-n) - k_{\text{off}}n + D\pdif{^2n}{s^2}.     \label{kinetic eq}
        \end{align}
        According to the Bell-Evans formula \cite{bell1978models,evans2007forces}, the mechanical force $f_b$ felt by a given linker will lower the energy barrier for it to unbind from the substrate, such that the off-rate increases exponentially with the force as $k_{\text{off}} = k_{\text{off}}^0 \text{exp}(f_b/f_0)$, where $k_{\text{off}}^0$ is the off-rate under zero force and $f_0$ is a molecular force scale for the linker to rupture with a typical order of several pN. 
        For simplicity, the on-rate $k_{\text{on}}$ is assumed to be a force-independent constant. 

        The mechanosensitivity of the adhesions allows us to capture the biphasic relation between the friction force and the retrograde velocity $v_r$ \cite{filippov2004friction,sabass2010modeling,li2010model,craig2015model}. 
        Here, we adopt a minimal mean-field approximation \cite{sens2020stick} where the average extension of a linker is approximated by the retrograde velocity $v_r$ times its average lifetime $1/k_{\text{off}}$. 
        Each linker is viewed as an elastic spring with spring constant $k_b$. 
        By Hooke's law, the force experienced by a single linker is given by $f_b = k_b v_r/k_{\text{off}}$.
        Hence, the retrograde velocity and the dimensionless off-rate $r = k_{\text{off}} / k_{\text{off}}^0$ can be related by $v_r = v_\beta r\log r$, where $v_\beta = k_{\text{off}}^0 f_0/k_b$ is a characteristic velocity scale related to the mechanosensitive unbinding process. 
        In addition to the friction generated by the mechanosensitive adhesions, viscous dissipation between the actin flow and the substrate contributes to a linear friction $\zeta_0 v_r$.\ 
        Assuming that protrusions as well as the retrograde flow are locally normal to the cell boundary \cite{lee1993principles,nickaeen2017free}, the total friction force is given by
        \begin{align}
            \mbf{F}_{\text{friction}} = \left(\zeta_0 v_r + \zeta_1 \frac{nv_r}{r}\right) \mbf{n}, 
        \end{align}
        where $\mbf{n}$ is a unit outward normal vector on the cell edge $\Gamma(t)$.

        To highlight the interplay between the mechanosensitive adhesions and the membrane tension, we simply treat the contractile force per unit length as a constant force pointing inwards along the normal direction, $\mbf{F}_{\text{contraction}} = -\sigma_c\mbf{n}$. 
        The time scale for the rebinding and unbinding of adhesions (seconds) \cite{chan2008traction} is much longer than the time scale for the membrane force to equilibrate (milliseconds) \cite{keren2008mechanism}, so the membrane tension, $\sigma_m$, is assumed to be spatially uniform along the cell boundary. The force per unit length integrated over the lamellipodium height is then given by $\mbf{F}_{\text{membrane}} = -2h\sigma_m H\mbf{n}$, where $H=1/h+\kappa$ is the total curvature given by the in-plane curvature $\kappa$ and the lamellipodium radius $h$ along the vertical direction to the substrate.
        
        We further assume that the membrane tension depends linearly on the total area $A(t)$ of the cell as it evolves in time, 
        \begin{align}
            \frac{\dd \sigma_m}{\dd t}(t) = k_\sigma \frac{\dd A}{\dd t}, 
        \end{align}
        where $k_\sigma$ is an effective stiffness that accounts for the extensibility of the membrane. 
        Finally, the shape of the cell evolves according to the kinematic boundary condition
        \begin{align}
            \pdif{\mbf{x}}{t}(\alpha,t) = (v_p - v_r)\mbf{n}, 
        \end{align}
        where the normal velocity is determined by the difference between the actin polymerization velocity $v_p$ and the retrograde velocity $v_r$. 
        Here, we treat the polymerization velocity as a constant along the cell edge following \cite{callan2008viscous,blanch2013spontaneous}, although it can be a function of the myosin density as hypothesized in some more detailed modeling approaches \cite{nickaeen2017free}. 

        \subsection*{Non-dimensionalization}
        We nondimensionalize the governing equations with the following scales. 
        The characteristic time scale is given by the off-rate under zero force load $1/k_{\text{off}}^0$. The mechanosensitive unbinding process provides a characteristic velocity scale $v_\beta$ as mentioned in the previous subsection. 
        The cell size is characterized by its average radius $R_0$. 
        The dimensionless variables are listed as follows:
        \begin{align}
            t^\ast = t k_{\text{off}}^0, \quad v_r^\ast = \frac{v_r}{v_\beta},\quad \mbf{x}^\ast = \frac{\mbf{x}}{R_0},\quad \sigma_m^\ast = \frac{2\sigma_m}{\zeta_0 v_\beta}. 
        \end{align}
        The dimensionless parameters governing the system can be classified into two groups, namely parameters that characterize the stick-slip dynamics of the adhesions and parameters that characterize the general cell properties. 
        We estimate their orders of magnitude from experiments and previous modeling approaches (see \textit{SI}), as summarized in Table \ref{dimensionless parameters}.\ 
        The first group contains the dimensionless relative adhesion strength $\zeta_1^\ast = {\zeta_1}/{\zeta_0}$ that compares the contribution of the adhesive friction (sticking) and the viscous friction (slipping), the dimensionless on-rate $r_{\text{on}} = k_{\text{on}}/k_{\text{off}}^0$, and the dimensionless diffusion coefficient $D^\ast = D /k_{\text{off}}^0 R_0^2$. 
        The second group contains the dimensionless lamellipodium height $h^\ast = h/R_0$, the dimensionless length $\epsilon = v_\beta/k_{\text{off}}^0 R_0$ that compares the length scale provided by the off-rate and the retrograde velocity with the cell size, the dimensionless effective stiffness $k_\sigma^\ast = 2k_\sigma R_0/\zeta_0 k_{\text{off}}^0$ that compares the cell stiffness to the adhesive friction, the dimensionless actomyosin contraction $\sigma_c^\ast = \sigma_c / \zeta_0 v_\beta$, and the dimensionless polymerization velocity $v_p^\ast = v_p/v_\beta$. 
      
        \begin{table}[t]
            \centering
            \caption{Range of dimensionless model parameters. See \textit{SI Appendix} for dimensional parameter values.}
            \begin{tabular}{lr}
            Dimensionless parameter         & Approximate value  \\
            \hline
                $\zeta_1^\ast = \zeta_1/\zeta_0$   & $400-600$ \\
                $r_{\text{on}}=k_{\text{on}}/k_{\text{off}}^0$ &  $10-60$ \\
                $D^\ast=D/k_{\text{off}}^0 R_0^2$ & $0.01-0.1$ \\
            \hline
                $\epsilon=v_\beta / k_{\text{off}}^0 R_0$ & $0.001$ \\
                $h^\ast = h/R_0$ & $0.01$ \\
                $\sigma_c^\ast = \sigma_c/\zeta_0 v_\beta$ & $100$\\
                $k_\sigma^\ast = 2k_\sigma R_0/\zeta_0 k_{\text{off}}^0$ & $1000$ \\
                $v_p^\ast=v_p/v_\beta$ & $200$\\
        \hline
        \end{tabular}
        \label{dimensionless parameters}
    \end{table}

    The dimensionless governing equations can be summarized as follows, where the stars have been dropped for simplicity: 
    \begin{align}
        & \pdif{\mbf{x}}{t}(\alpha,t) = \epsilon(v_p - r\log r)\mbf{n}, \label{shape} \\
        & \pdif{n}{t}(\alpha,t) = r_{\text{on}}(1-n) - rn + D \pdif{^2n}{s^2}, \label{kinetic}\\
        & \frac{\dd \sigma_m}{\dd t}(t) = k_\sigma \frac{\dd A}{\dd t}(t),\label{elasticity}\\ 
        & \sigma_c + \sigma_m(1 + \kappa h) = r\log r + \zeta_1 n\log r. \label{force balance}
    \end{align}

    \subsection*{Numerical implementation}

    To describe the geometry, the tangent angle, $\theta$, and the arclength derivative, $s_\alpha$, are introduced as independent variables instead of $\mbf{x}$. 
    By specifying the arclength derivative as $s_\alpha = L/2\pi$, the mesh points are kept equally spaced in arclength at every time step. 
    The derivatives with respect to $s$ and $\alpha$ can be exchanged through $\pl_\alpha = s_\alpha \pl_s$: this enables us to apply a finite difference scheme in the fixed $\alpha$-parametric domain, to which the curve is mapped as a circle and uniformly discretized in $\alpha$ as $\alpha_i = 2\pi(i-1)/N,\;i = 1,\cdots,N+1$. The method was validated by comparison to the linear stability results at short times (see Fig.~S2 of the \textit{SI Appendix}), and by comparison with a different scheme based on spline interpolation; excellent agreement was found between the two schemes, with the $\theta$--$L$ method providing enhanced computational speed.

    {All simulations start from a circular configuration with a uniform fraction of bound linkers, which is perturbed} initially, {along with} other corresponding physical quantities, using the first $100$ Fourier modes with amplitudes varying randomly from $-10^{-4}$ to $10^{-4}$. 
    Given the {geometric and physical variables} at time step $t^n$, we update their values at $t^{n+1}$ by the following steps:
    (1) Update the shape of the curve $\mbf{x}$ with the $\theta$--$L$ formulation using an explicit Euler scheme. Compute other geometric quantities such as the normal vector and the curvature.\ 
    (2) Update the membrane tension $\sigma_m$ with the explicit Euler scheme. 
    (3) Update the fraction of bound linkers $n$ with a Crank-Nicholson scheme. 
    (4) Update the off-rate by solving the force balance \eqref{force balance} iteratively using Newton's method and obtain the normal velocity. 
    The spatial derivatives are discretized by central finite difference and the integrals are computed by the trapezoidal rule with end-correction. 
    The number of grid points is taken as $N = 2000$ and the time step is set to $\Delta t = 10^{-3}$. 

    \section*{Results}
        \subsection*{Initiation of cell motion through a stick-slip instability}
        We first analyze the linear stability of the model to uncover a mechanism for cell spontaneous symmetry breaking and motility initiation through an instability arising from the stick-slip dynamics of the mechanosensitive adhesions. 
        A similar instability was discussed in \cite{sens2020stick} for the 1D case as a ``stick-slip instability'', which gave rise to persistent oscillations between protrusion and retraction phases by switching between sticking and slipping states. 
        Here, we show that, when cell shape change is considered, this instability still exists but distinct modes of deformation are subject to distinct instability criteria due to the spatial effects.

        We take the base state for the analysis to be a stationary circle with radius $\bar{R} = 1$ where the retrograde velocity balances with the protrusion velocity $v_p = \bar{r}\log \bar{r}$ everywhere, and the fraction of bound linkers is uniformly distributed along the edge with the steady-state value $\bar{n} = r_{\text{on}}/(r_{\text{on}} + \bar{r})$. Henceforth, overbars are used to denote base-state variables.  
        From the force balance \eqref{force balance}, the base state surface tension then satisfies
        \begin{align}
            \bar{\sigma}_{m} = \frac{1}{1+h/\bar{R}}(\bar{r}\log \bar{r} + \zeta_1 \bar{n} \log \bar{r} - \sigma_c). 
        \end{align}
        We consider small perturbations around the base state with the ansatz $\phi = \bar{\phi} + \delta \phi = \bar{\phi} + \sum_k \phi_k \text{exp}(\ii k\theta + \lambda_k t)$. 
        Here $\bar{\phi}$ represents the base-state value of variable $\phi$, and $\phi_k$ and $\lambda_k$ denote the initial magnitude of perturbation and corresponding dimensionless growth rate of the $k^\mathrm{th}$ normal mode, respectively.  
        Note that in our model, the surface tension is assumed to be uniform along the cell edge, so perturbations of the surface tension with modes $k\ne 0$ are set to zero. 
        Inserting the ansatz into the governing equations and linearizing the system for small perturbations yields eigenvalue problems for the growth rates $\lambda_k$. For $k=0$,
        \begin{align}
            &  \Big(1+\log \bar{r} + \zeta_1\frac{\bar{n}}{\bar{r}}\Big)\lambda_0^2 
            + 
             \Big\{\Big(1+\log \bar{r} + \zeta_1\frac{\bar{n}}{\bar{r}}\Big)(r_{\text{on}} + \bar{r})  \notag \\
            & +\!\Big[\Big(1+\frac{h}{\bar{R}}\Big)2\pi \bar{R} k_\sigma \!-\! \frac{\epsilon h\bar{\sigma}_{m}}{\bar{R}^2}\Big](1+\log \bar{r}) \!-\! \zeta_1 \bar{n}\log \bar{r}\Big\}\lambda_0 \notag \\
            & +\! \Big[\Big(1+\frac{h}{\bar{R}}\Big)2\pi \bar{R} k_\sigma \!-\! \frac{\epsilon h\bar{\sigma}_{m}}{\bar{R}^2}\Big](1\!+\!\log \bar{r})(r_{\text{on}} \!+\! \bar{r}) = 0,
        \end{align}
        and for $k\neq 0$,
        \begin{align}
            &   \Big(1\!+\!\log \bar{r} \!+\! \zeta_1\frac{\bar{n}}{\bar{r}}\Big)\lambda_k^2 \!+\! 
             \Big[ \Big(1\!+\!\log \bar{r} \!+\! \zeta_1\frac{\bar{n}}{\bar{r}}\Big)\Big(r_{\text{on}}\!+\!\bar{r} \!+\! \frac{Dk^2}{\bar{R}^2}\Big)  \notag \\
            & +\frac{\epsilon h\bar{\sigma}_{m}}{\bar{R}^2}(k^2-1)(1+\log \bar{r}) - \zeta_1 \bar{n}\log \bar{r}\Big]\lambda_k \notag \\
            & + \frac{\epsilon h\bar{\sigma}_{m}}{\bar{R}^2}(k^2-1)(1+\log \bar{r})\Big(r_{\text{on}}+\bar{r}+\frac{Dk^2}{\bar{R}^2}\Big) = 0. 
        \end{align}
        The governing equations are nonlinearly coupled and, consequently, they provide constraints on the initial perturbations (see the \textit{SI} for details).
        These constraints satisfy a quadratic equation for the perturbation amplitudes with two possible conjugate imaginary solutions, corresponding to two independent modes of perturbation, analogous to the one-dimensional case \cite{sens2020stick} where the two independent modes are the symmetric and the anti-symmetric modes. 
        
\begin{figure}[t]
            \centering
            \includegraphics[width=8.5cm]{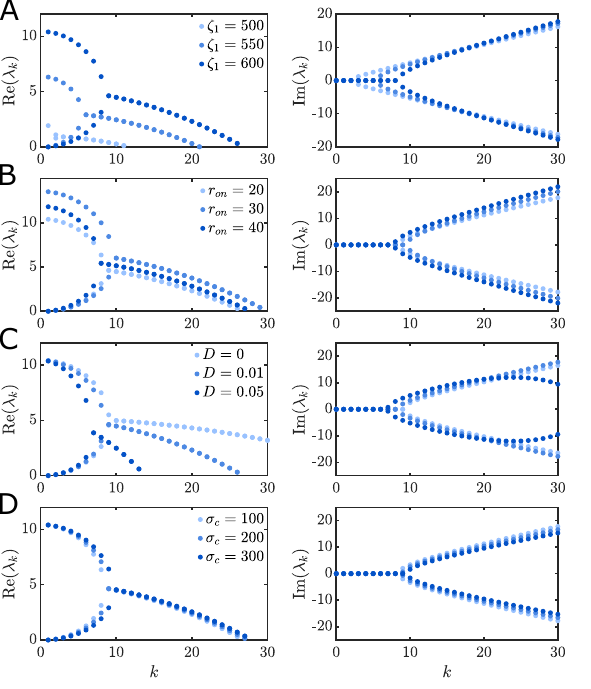}
            \caption{Dispersion relations showing the real and imaginary parts of the growth rate $\lambda_k$ as functions of wavenumber $k$ for various choices of the parameters related to adhesion kinetics: (\textit{A}) growth rates under different relative adhesion strengths $\zeta_1$ with $r_\text{on} = 20,\; D = 0.01,\;\sigma_c=100$; (\textit{B}) growth rates under different on-rates $r_\text{on}$ with $\zeta_1 = 600,\; D = 0.01,\;\sigma_c=100$; (\textit{C}) growth rates under different diffusion coefficients $D$ with $\zeta_1 = 600,\; r_\text{on} = 20,\;\sigma_c=100$; (\textit{D}) growth rates under different contraction strengths $\sigma_c$ with $\zeta_1 = 600,\; r_\text{on} = 20,\;D=0.01$. In all cases, the parameters related to the general cell properties are fixed as $\epsilon = 0.001,\;h = 0.01,\;k_\sigma = 1000,\; r_p = 50$. }
            \label{dispersion relations stick-slip}
        \end{figure}

        The various Fourier modes represent different modes of deformation. Typical dispersion relations for parameter choices relevant to physiological conditions values are plotted in Fig.~\ref{dispersion relations stick-slip}. 
        We find that the precise choice of parameters does not affect the qualitative behavior of the dispersion relation, which can be summarized as follows.\ 
        The mode $k=0$ describes a spatially homogeneous perturbation and corresponds to the global 
        dilation or contraction of the cell. 
        This is the only mode to be affected by the effective membrane stiffness $k_\sigma$, and we find it to be always linearly stable at physiological values of the stiffness ($k_\sigma \sim 10^3$). 
        Thus, membrane elasticity always acts to maintain the cell area at its base value in the linear regime. 
        The mode $k=1$ is the only mode that is not center-symmetric and captures translational motion of the center of mass; its growth rate is found to be purely real regardless of the choice of model parameters. 
        Therefore, when this mode becomes unstable, the corresponding perturbation is amplified and leads to a symmetry breaking in space that singles out a certain direction to initiate locomotion. 
        Subsequent modes with $k>1$ describe various shape deformations with increasingly shorter wavelengths. At relatively low wavenumbers, the dispersion relation typically displays two real positive growth rates. As $k$ increases, these give way to two complex conjugate growth rates, suggesting that there can exist oscillations and traveling waves propagating along the edge, known as ``stick-slip'' waves, which are similar to the lateral waves predicted along a flat edge \cite{sens2020stick}.\ Finally, all modes beyond a certain wavenumber become stable, indicating that high-frequency perturbations will decay.

        The dependence of the growth rates on the relevant system parameters can also be gleaned from Fig.~\ref{dispersion relations stick-slip}. We first note that a Hopf bifurcation can occur for modes with high wavenumbers as the parameters vary, where the real growth rates become growth rates with conjugate imaginary parts, similar to the 1D case \cite{sens2020stick}. 
        We focus on the role of the three parameters that characterize adhesion kinetics. 
        The relative adhesion strength $\zeta_1$ and the on-rate $r_\text{on}$ are found to affect both the magnitude of the unstable growth rates as well as the number of unstable modes, while the effect of the diffusion coefficient $D$ is to damp high order modes. 
        As $\zeta_1$ increases, both the magnitude of the positive real growth rates and the number of unstable modes increase. 
        The effect of the on-rate $r_\text{on}$, on the other hand, is nonmonotonic and exhibits a biphasic behavior.\ 
        When the on-rate increases, the magnitudes of unstable growth rates first increase and then decrease. 
        This suggests the possibility of a stick-slip instability similar to the 1D case \cite{sens2020stick}, and provides a mechanism for the cell to oscillate between protruding (sticking) and retracting (slipping) states at the two distinct ends. 
        Here, in two dimensions, certain modes of perturbation are amplified by the mechanosensitive nature of the adhesion clusters and their coupling with the retrograde flow, leading to the local adhesion sites switching between sticking and slipping motions in distinct regions. 
        To make the biphasic relationship between friction and retrograde flow possible, the relative adhesion strength needs to be strong enough for the adhesive friction to play a role compared to the linear friction. 
        Moreover, the rebinding rate cannot be either too small or too large, since in both cases the adhesive linkers either dissociate or rebind to the substrate too quickly. As a result, there is no possible stick-slip transition and the system remains linearly stable. 
        That optimal locomotion efficiency occurs for intermediate adhesions was demonstrated in experiments by varying the concentration of integrins and blocking integrins \cite{palecek1997integrin}. 
        
        Our model assumed a uniform contractile force along the cell edge as a way to single out the role of mechanosensitive adhesions in driving symmetry breaking. The effect of actomyosin contraction on stability is shown in Fig.~\ref{dispersion relations stick-slip}$D$, where it is found to have a weak but destabilizing effect on the system. This finding is qualitatively consistent with previous contraction-driven cell motility models and with  various experimental observations \cite{recho2013contraction, barnhart2015balance, safsten2022asymptotic}. 
        
        \subsection*{Instability of mode $k=1$ is responsible for the formation of the front and rear in the early stage of locomotion}
\begin{figure*}[t]
                \centering
                \includegraphics[width=0.99\textwidth]{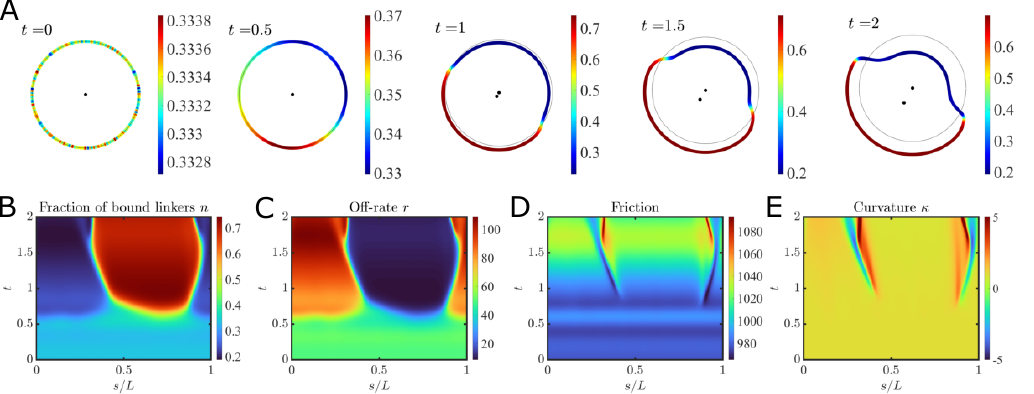}
                \caption{Spatiotemporal evolution of a fan-shaped cell during the initiation of motion ($t=0$ to $t=2$) from a nonlinear numerical simulation. (\textit{A}) Time evolution of the cell shape, where the edge is colored by the fraction of bound linkers. The two black dots show the initial and current center-of-mass positions. (\textit{B})-(\textit{E}) Kymographs of the fraction of bound linkers $n$,  off-rate $r$,  total friction $r\log r + \zeta_1 n\log r$, and  local curvature $\kappa$, as functions of normalized arclength $s/L$ and time $t$. Parameters values: $\epsilon = 0.001,\;h = 0.01,\;k_\sigma = 1000,\;\sigma_c = 100,\; r_p = 50,\;r_\text{on}=25,\;\zeta = 600,\;D=0.05$. }
                \label{spatial-temporal short time}
            \end{figure*}

        As already mentioned above, the instability of the $k=1$ mode provides a mechanism for the self-polarization of the cells and initiation of motility. 
        This is further demonstrated in our numerical simulations. 
        A typical temporal evolution of a cell shape at short times is illustrated in Fig.~\ref{spatial-temporal short time}$A$, where the cell edge is colored by the fraction of bound linkers. 
        Starting from a circular shape and a randomly perturbed initial distribution of bound linkers at $t=0$, high-frequency fluctuations are found to decay very rapidly ($t=0.5$) as predicted by the stability analysis. 
        The $k=1$ mode, which has the largest growth rate, grows simultaneously, resulting in the formation of a potential cell front where more linkers are bound to the substrate and of a potential rear where fewer linkers are bound ($t=1$). 
        The polymerization velocity then exceeds the retrograde velocity in the front while the opposite occurs at the rear, leading to expansion in the front, shrinkage at the rear, and translocation of the center of mass ($t=1.5$). 
        The front further keeps protruding while the rear keeps retracting, and
        the cell ultimately evolves to a fan-like shape with a smooth leading edge ($t=2$) very similar to the characteristic shape of coherent keratocytes \cite{barnhart2010bipedal}. 

        The spatiotemporal evolution of the fraction of bound linkers, off-rate, friction force and curvature during these early stages are plotted as kymographs in Fig.~\ref{spatial-temporal short time}$B$--$E$. 
        As a result of the growth of the $k=1$ mode, the retrograde velocity is low in the front of the cell and high at the rear. 
        Consequently, strong adhesion develops in the front where the off-rate is low and hence most linkers are bound to the substrate. Meanwhile the off-rate is high with most linkers unbound along regions at the back and sides, in agreement with the traction stress distribution revealed by experiments in migrating cells \cite{fournier2010force}. 
        Comparing the spatiotemporal distribution of the friction and the curvature (Fig.~\ref{spatial-temporal short time}$D$ and $E$) shows that regions with high friction coincide with regions with high curvature as required by the force balance equation. 
        This is also consistent with experiments \cite{fournier2010force} where traction stress concentrates at the sides. 
        Note that unlike many previous modeling approaches \cite{barnhart2010bipedal,lee2020modeling} where the breaking of front-rear symmetry and adhesion distribution were treated as inputs to trigger the initiation of motion and turning, our model provides a route for this distribution of adhesions to emerge spontaneously through the instability of the $k=1$ mode and to become further amplified by the feedback loop between mechanosensitive adhesions, membrane tension, and geometry. 

        \begin{figure*}[t]
        \centering
        \includegraphics[width=0.96\textwidth]{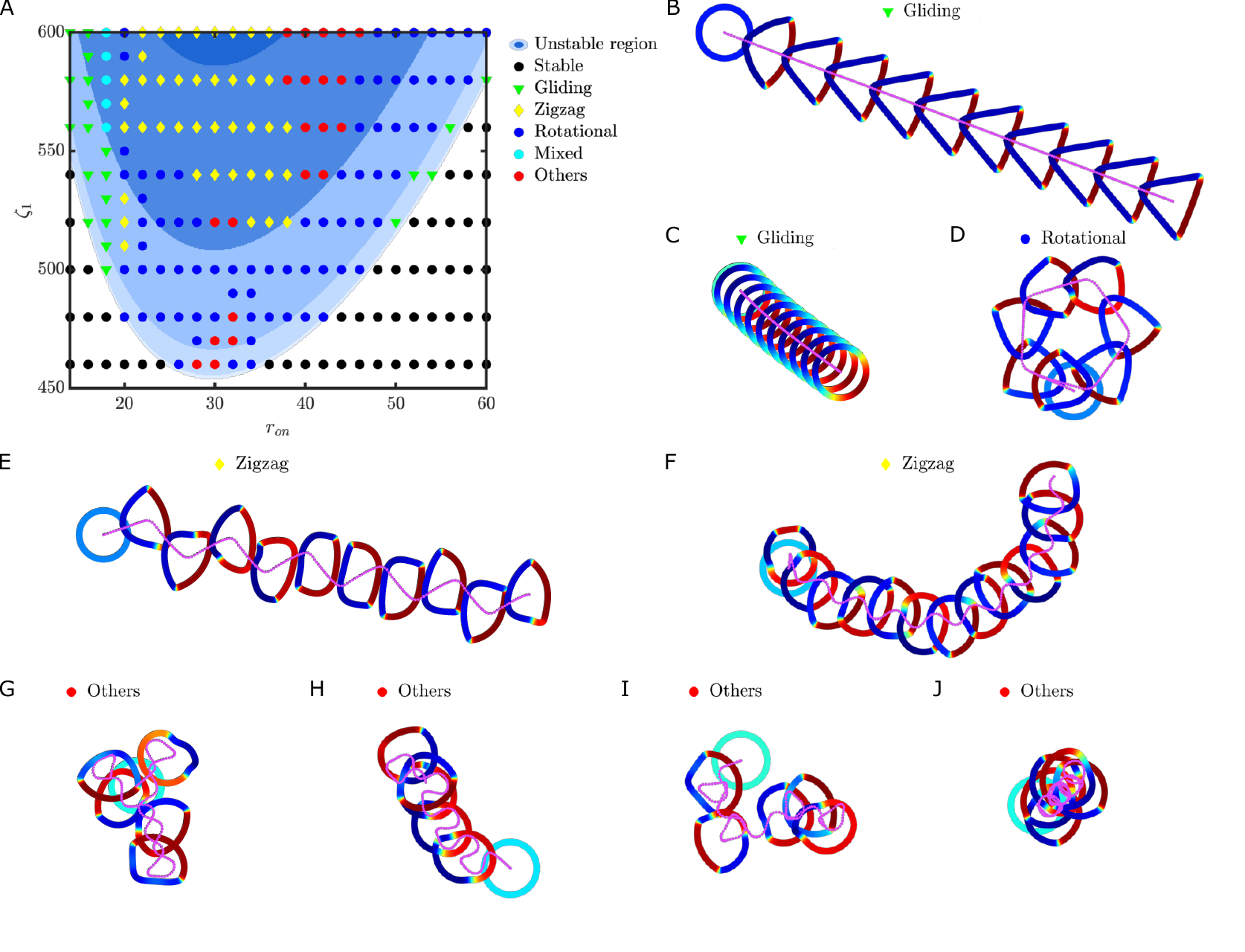}
        \vspace{-0.1cm}
        \caption{(\textit{A}) Stability diagram and motility modes in the $(r_{\text{on}},\zeta_1)$ parameter space: blue shadings show regions of instability for modes $k=1,\;2,\;5,\;10,\;15$ (from light to dark blue), and symbols indicate the motility modes observed in nonlinear simulations, whose dynamics are illustrated in the remaining panels. Five simulations were performed for each data point. Points labeled as `mixed' display multiple types of dynamics depending on the initial condition. Points labeled as `others' display irregular dynamics that are not easily categorized.  (\textit{B}) Gliding mode, $(r_{\text{on}},\alpha_1) = (16,600)$, $\Delta t= 10$. (\textit{C}) Gliding mode, $(r_{\text{on}},\alpha_1) = (56,560)$, $\Delta t= 10$. (\textit{D}) Rotational mode, $(r_{\text{on}},\alpha_1) = (24,540)$, $\Delta t= 10$. (\textit{E}) Zigzag mode, $(r_{\text{on}},\alpha_1) = (26,600)$, $\Delta t= 10$. (\textit{F}) Zigzag mode, $(r_{\text{on}},\alpha_1) = (36,560)$, $\Delta t= 15$. (\textit{G}) Others, $(r_{\text{on}},\alpha_1) = (38,580)$, $\Delta t= 20$. (\textit{H}) Others, $(r_{\text{on}},\alpha_1) = (40,540)$, $\Delta t= 20$. (\textit{I}) Others, $(r_{\text{on}},\alpha_1) = (40,600)$, $\Delta t= 20$. (\textit{J}) Others, $(r_{\text{on}},\alpha_1) = (44,560)$, $\Delta t= 20$. In all cases, the other parameters were set to $\epsilon = 0.001,\;h=0.01,\;k_\sigma=1000,\;\sigma_c=100,\;r_p=50,\;D=0.05$. The cells contours are colored by the fraction of bound linkers $n$, and the purples curves show the center-of-mass trajectories in each case. See Movies S1--S9 in the \textit{SI Appendix} for videos of the dynamics.}\vspace{-0.1cm}
        \label{stability diagram and motility modes}
    \end{figure*}

    \subsection*{The stability diagram predicts various motility modes}
Next, we turn our focus to long-time dynamics and the effect of adhesion kinetic parameters. While the initiation of locomotion through the instability of the $k=1$ mode is generic, various cell motility modes are observed at long times depending on the relative adhesion strength $\zeta_1$ and on-rate $r_{\text{on}}$. We categorize them in a phase diagram in Fig.~\ref{stability diagram and motility modes}$A$, in which the blue shaded regions highlight the linearly unstable regimes for the Fourier modes with wavenumbers $k = 1,\;2,\;5,\;10,\;15$, while the dots denote the long-time motility modes.\ 
Each dot represents five simulations with different random initial perturbations.\ A systematic exploration of the parameter space allowed us to classify the motility modes by the following criteria: (i)~if the cell moves in a straight line with no turns over the entire  simulation time range ($t=100$), we call it a ``gliding'' mode (Fig.~\ref{stability diagram and motility modes}$B$ and $C$); (ii)~if the cell turns only towards one direction, we call it a ``rotational'' mode (Fig.~\ref{stability diagram and motility modes}$D$); (iii)~if the cell turns alternatively between left and right resulting in a lateral oscillation, we call it a ``zigzag'' mode (Fig.~\ref{stability diagram and motility modes}$E$ and $F$); (iv)~for some combinations of parameters and for different initial conditions, we observed ``mixed'' modes where the cell can exhibit different types of motions including the three cases mentioned above; (v)~finally, there exist more complicated types of motion that do not fit into any of the cases above and that we label as ``others'' (Fig.~\ref{stability diagram and motility modes}$G$--$J$). 
Most of the motility modes described here have been observed in experiments as well as in previous models: the gliding motion is well known in fast-moving fish keratocytes \cite{verkhovsky1999self}; the zigzag motion is analogous to the ``bipedal'' motion of oscillating keratocytes \cite{barnhart2010bipedal,loosley2012stick}; and spontaneous turning has also been observed in keratocytes even in the absence of external cues \cite{oliver1999separation,allen2020cell}. 

\begin{figure}[t]
    \centering
    \includegraphics[width=8.5cm]{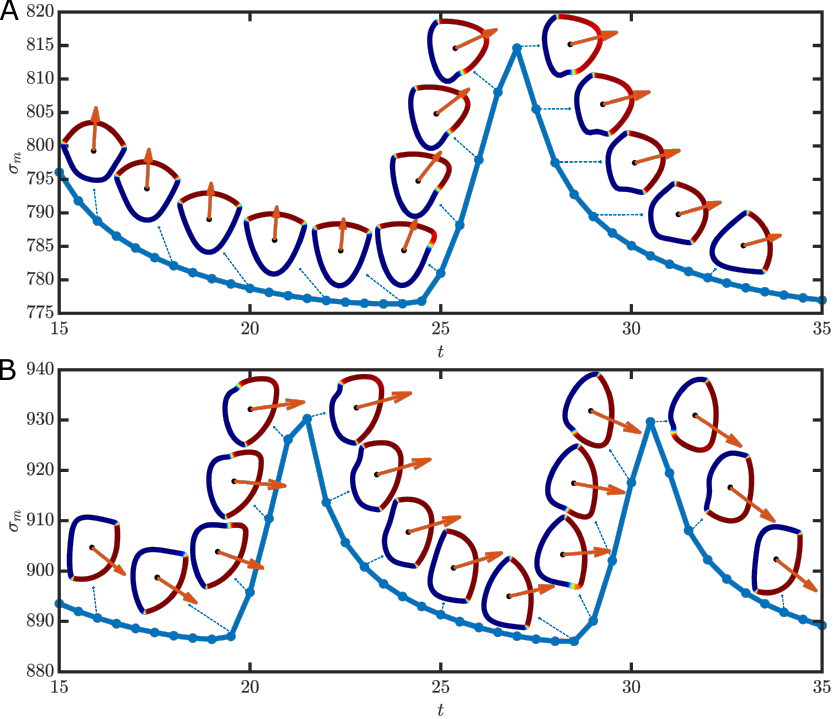}
    \caption{Temporal evolution of the membrane tension and corresponding cell shapes and center-of-mass velocities as functions of time during one period of oscillation for the (\textit{A}) rotational mode with  $(r_{\text{on}},\alpha_1) = (24,540)$, and (\textit{B}) zigzag mode with $(r_{\text{on}},\alpha_1) = (26,600)$. The cell shapes are colored by the local fraction of bound linkers $n$, and the orange arrows show the instantaneous center-of-mass velocity.
    In both cases, other parameters are fixed as $\epsilon = 0.001,\;h=0.01,\;k_\sigma=1000,\;\sigma_c=100,\;r_p=50,\;D=0.05$.}
    \label{fig: tension and shape}
\end{figure}

Although the dynamics at long times are essentially nonlinear, Fig.~\ref{stability diagram and motility modes}$A$ shows a clear correlation between the stability diagram and the motility modes. 
In regions of the parameter space with fewer unstable modes, where the on-rate is either relatively low or high, the motion tends to be uniaxial with the cell remaining polarized with left-right symmetry. 
More complex dynamics arise for intermediate on-rates, which we attribute to the stick-slip instability. 
Rotational and zigzag modes arise in this regime, and are characterized by broken left-right symmetry in addition to front-rear asymmetry, resulting in quasi-periodic turns in the cell trajectory. 
We take a closer look at these turns in Fig.~\ref{fig: tension and shape}\textit{A} and \textit{B}, where we plot the temporal evolution of the membrane tension together with the corresponding shapes and center of mass velocities within one oscillation period for the rotational and zigzag modes, respectively. 
At the initiation of a turn, the distribution of bound linkers is first observed to become asymmetric, with an increase in the fraction of bound linkers on one side of the cell. 
This results in enhanced sticking on that side along with slipping on the opposite side, allowing the cell to turn in that direction, and this mechanism is reminiscent of the ``sticking wave'' predicted in the 1D flat case \cite{sens2020stick}. 
As the cell shape becomes asymmetric and the cell rotates, the membrane tension is found to increase and reaches a peak value before relaxing again in the later stage of the turn. 
In the zigzag case, the high adhesion region alternatively switches between the left and right sides of the cell, while for the rotational cases the direction in which the high adhesion regions form remains fixed, with equal probabilities for a cell to turn clockwise or counterclockwise. The Hopf bifurcation of higher-order modes may be responsible for this transition from uniaxial translation to oscillatory motion. 
Moreover, we find that there is a strong correlation between cell shapes and motility modes. 
For the gliding cases, the cell adopts either a triangular-like shape (for small on-rates, Fig.~\ref{stability diagram and motility modes}\textit{B}) or a nearly circular shape (for large on-rates, Fig.~\ref{stability diagram and motility modes}\textit{C}), while for the zigzag and rotational motions, the cells are more fan-shaped with curved fronts (Fig.~\ref{stability diagram and motility modes}\textit{D}--\textit{F}). 
Another key observation is that, regardless of the randomness in the initial perturbations, the emerging shapes corresponding to a give set of parameters are robust; we further elaborate on this point below.

\subsection*{Robustness of motility modes} 
\begin{figure}[t]
            \centering
            \includegraphics[width=8.5cm]{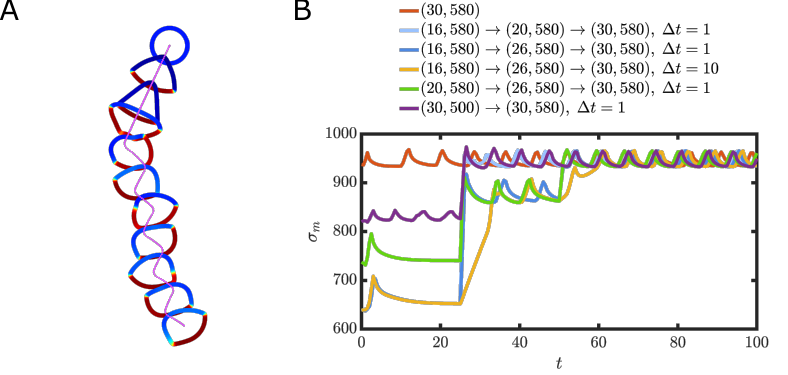}
            \caption{Transition between different motility modes. (\textit{A}) Snapshots of a gliding cell switching into a zigzag mode as the adhesion parameters $(r_\text{on},\alpha_1)$ abruptly change from $(16,580)$ to $(30,580)$ at $t=25$. Other parameters are fixed as $\epsilon = 0.001,\;h = 0.01,\;k_\sigma = 1000,\; r_p = 50,\;D=0.05$.  (\textit{B}) Temporal evolution of the membrane tension as the parameters $(r_\text{on},\alpha_1)$ are varied to cause changes in motility modes. Here, $\Delta t$ denotes the duration of the time interval over which the parameters are gradually varied. }
            \label{tansition}
        \end{figure}

The observations above suggest that the feedback loop between mechanosensitive adhesions and membrane tension uniquely determines the emergent motility mode for a given set of system parameters, irrespective of the initial condition or history of the system.\ 
To further demonstrate the robustness of these modes, we vary the parameters during a simulation to analyze transitions between modes.\ 
A typical transition is shown in Fig.~\ref{tansition}\textit{A}: starting from a cell performing a gliding motion, we abruptly vary the value of $r_{\text{on}}$ from $16$ to $30$ at $t_0=25$, causing it to switch to a zigzag mode.  
As shown in the snapshots, the cell glides smoothly at first, and, after the change in $r_{\text{on}}$, its shape quickly adjust and start undergoing zigzags. 
The oscillation frequency and cell morphology after the transition are similar to the corresponding zigzag motions with $r_{\text{on}}=30$ starting from a randomly perturbed initial condition. 
The time instant $t_0$ at which we start altering the parameters, the period of time $\Delta t$ within which we gradually alter the parameters, and the intermediate states do not significantly affect the zigzag dynamics (Fig.~\ref{tansition}\textit{B}). 
To quantitatively compare the resulting zigzag modes emerging through different routes, we plot the temporal evolution of the membrane tension in Fig.~\ref{tansition}\textit{B} for different choices of $t_0$ and $\Delta t$. 
After increasing the on-rate and the adhesion strength, the cell expands, leading to an increase in the membrane tension towards the tension value of the new state, regardless of the history of states. 
This is in agreement with the one-dimensional case \cite{sens2020stick} as well as experiments \cite{lieber2013membrane} where the membrane tension in motile cells is determined by the adhesion strength and cytoskeletal forces, and increases as cell-substrate adhesion strengthens---in our model, as either the on-rate and relative adhesion strength increases. 

Similar transitions are also observed when we change parameters between other modes (see \textit{SI Appendix}, Movies S10--S12). 
In all cases, after changing the parameters, the cell first goes through a transient state and then soon evolves into the motility mode selected by the new set of parameters in the phase diagram of Fig.~\ref{stability diagram and motility modes}. 
This further suggests that these motility modes are attractors of the dynamical system, with the model solutions either approaching fixed points or stable limit cycles in phase space. 

        \subsection*{Effect of adhesion parameters} 
\begin{figure*}[t]
            \centering
            \includegraphics[width=0.98\textwidth]{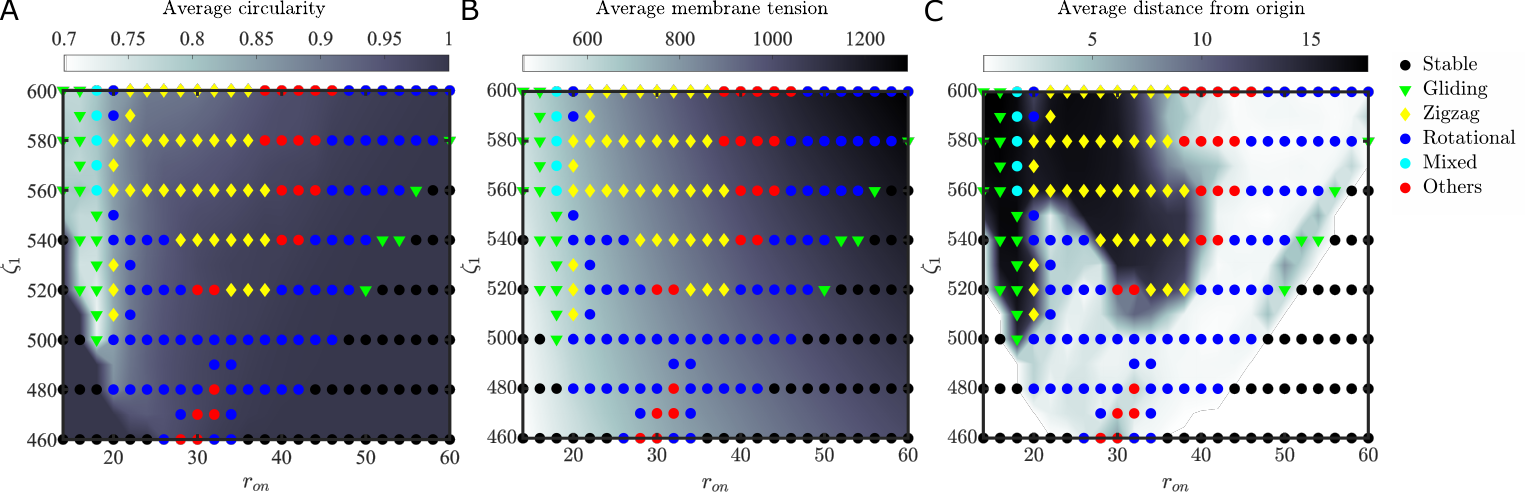}
            \caption{Phase diagrams in the $(r_{\text{on}},\zeta_1)$ parameter space showing color maps of the time-averaged (\textit{A}) cell circularity $4\pi A/ L^2$, (\textit{B}) membrane tension $\sigma_m$, and (\textit{C}) average distance of the center of mass from the origin $x_c$. The diagrams show averages over five simulations of duration $T=95$. Symbols indicate the corresponding motility modes previous identified in Fig.~\ref{stability diagram and motility modes}. 
            Other parameters are fixed as $\epsilon = 0.001,\;h=0.01,\;k_\sigma=1000,\;\sigma_c=100,\;r_p=50,\;D=0.05$.} 
            \label{phase diagrams}
        \end{figure*}

        Finally, we investigate the effect of varying adhesion kinetic parameters on cell geometry, mechanics and locomotion. 
        For various combinations of $r_{on}$ and $\zeta_1$, we calculate the time-average circularity $4\pi A/ L^2$ (with a maximum value of 1 corresponding to a perfect circle), the average membrane tension, and the average distance of the center of mass from the origin $x_c=\sum_{i=1}^M |\mbf{x}_c^{(i)} - \mbf{x}_c^{(0)}|/M$, and plot them as phase diagrams in Fig.~\ref{phase diagrams} where the motility modes are also labeled. Note that the precise boundary between each mode changes slightly as we vary the diffusion coefficient $D$, but the qualitative behavior remains unchanged. 
        For the gliding modes with low on-rates, the cell adopts a nearly triangular shape with low circularity, with only a weak dependence on $\zeta_1$. 
        As the on-rate increases, the shapes all become more circular regardless of motility mode as seen in Fig.~\ref{phase diagrams}\textit{A}. 
        A possible explanation for this behavior is that, as the on-rate increases, the difference in the fraction of bound linkers between the front and the rear decreases, and therefore the difference in the sticking and slipping velocity also decreases, resulting in less deformed shapes. 
        The average membrane tension shows little correlation with the various motility modes in Fig.~\ref{phase diagrams}\textit{B}, and is mainly determined by the on-rate and the relative adhesion strength, consistent with the discussion above \cite{sens2020stick, lieber2013membrane}. 
        
        The ability of a cell to explore space depends strongly on its motility mode as illustrated in Fig.~\ref{phase diagrams}\textit{C}, where we show the phase diagram for the average distance traveled by the cell. 
        The gliding and the zigzag modes are more unidirectional and thus allow the cell to travel for a longer distance than in the rotational mode. 
        In the case of zigzag trajectories, the left-right oscillations can be accompanied by a circular motion when the relative adhesion strength and the on-rate increase, leading to a decrease in the distance traveled. Ultimately, the zigzag mode gives way to trajectories labeled as ``others'' in Fig.~\ref{phase diagrams}\textit{C}, which are characterized by irregular turns and may result from the superposition of multiple oscillatory periods.

        \section*{Discussion}
        
        In this work, we have extended the one-dimensional model coupling actin polymerization, adhesion, membrane tension, and shape change described in \cite{sens2020stick} to two dimensions to elucidate the role of the coupling between tension and adhesion in the determination of cell shape, the initiation of migration, and the selection of motility modes. 
        With this minimal model, we showed that motile cells can display rich dynamical behaviors by relying on a relatively simple set of physical mechanisms and couplings. 
        We first performed a linear stability analysis and demonstrated that the $k=1$ mode, describing translocation of the center of mass, is always the most unstable and is responsible for the cell to obtain its polarity and identify a direction in space for migration by spontaneous symmetry breaking. 
        We then conducted nonlinear numerical simulations, and showed that, driven by a constant actin polymerization rate, the model was able to capture various motility modes commonly seen in experiments, such as unidirectional gliding, bipedal motion and turning. 
        We identified the nonlinear coupling between stochastic adhesion and membrane tension as the key mechanism involved in the selection of motility modes. Specifically, the relative adhesion strength and on-rate were shown to govern membrane tension, which in turns connects spatially distributed adhesions and thus couples the retrograde flow with shape change and adhesion kinetics. 
        Certain fluctuation modes are amplified by this feedback loop, resulting in directional motion and oscillatory behaviors. 
        
\begin{figure}[t]
            \centering
            \includegraphics[width=8.5cm]{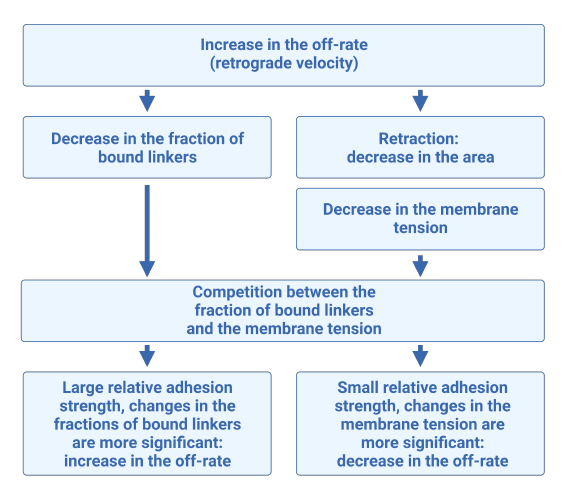}
            \caption{A simple mechanical feedback loop between membrane tension and adhesions determines cell shape and direction of motility.} 
            \label{fig:feedback loop}
        \end{figure}

        The basic physical mechanisms involved in this process are summarized in Fig. \ref{fig:feedback loop}. A local increase in the off-rate $r$ can have two distinct effects: a decrease in the fraction of bound linkers $n$ and a local retraction of the cell edge. 
        The latter will lead to a global decrease in cell area and, consequently, in a decrease in membrane tension due to the membrane elasticity.
        The force balance \eqref{force balance} dictates that the decrease in $n$ will cause an increase the off-rate $r$, whereas the decrease in membrane tension will cause $r$ to decrease. 
        Therefore, there is a competition between the membrane tension and the fraction of bound linkers in the regulation of the off-rate, tuned by the relative adhesion strength $\zeta_1$. 
        When $\zeta_1$ is large, the effect of $n$ dominates resulting in a positive feedback on $r$, rendering the system unstable and driving the onset of motility.
        On the other hand, when $\zeta_1$ is small, the situation is the opposite and the system is stable.  
        These predictions are borne out by the results of our stability analysis and numerical simulations.
        
        Our model predictions are consistent with experimental observations of symmetry breaking and cell motility modes. 
        Using experimental measurements and mechanical models, Barnhart \textit{et al.} showed that two feedback loops---one between actin flow and adhesions and the other between actin flow and myosins---are required to initiate motility in fish keratocytes \cite{barnhart2015balance}.
        Similarly, using a phase-field approach, Shao \textit{et al.} \cite{shao2012coupling} showed that coupling between adhesions, actin flow, and myosin contraction is required to recapitulate different experimental observations of keratocyte shape change. 
        In the present work, we showed that the feedback between adhesion and membrane tension, for constant actin flow, is sufficient to capture both the spontaneous initiation of motility by symmetry breaking and the emergence of complex motility modes.
        An interesting and as yet unexplained experimental observation that our model may help shed light on is that, at low temperatures, the trajectories of fast-moving keratocytes tend to be more unidirectional, while their motion is more circular and less persistent at high temperatures \cite{mogilner2020experiment}. 
        Previous studies have proposed a ``steering wheel'' mechanism \cite{allen2020cell,lee2020modeling} whereby the turning of a cell is caused by asymmetrically distributed adhesion sites at the rear, but the relationship between the turning and temperature is unclear. 
        This effect can be explained by the molecular clutch approach for adhesions in our model. 
        According to Bell's theory \cite{bell1978models}, the reaction rate increases with temperature, so binding and unbinding processes should be more active at high temperatures. By this effect, an increase in temperature drives an increase in the on-rate and relative adhesion strength, and thus drives the transition from gliding to zigzag, rotational, and other modes.

        While our model captures many experimental observations and makes testable predictions on the interaction between membrane tension and adhesion, it has certain limitations.
        In particular, it does not account for the complex molecular machinery underlying actomyosin contraction  or associated signaling pathways and only contains a single mechanical feedback loop. 
        We also note that even though we obtain fan-shaped cells in the early stages of motion, the shapes ultimately become triangular for gliding cells, which could be a result of the constant actin polymerization velocity assumed in our model and of ignoring actin remodeling events \cite{Xiong2010,rangamani2011,mogilner2020experiment,nickaeen2017free}.
        The contraction generated by the distinct distribution of myosin motors in crawling cells is also neglected in our model. 
        In some studies, contraction alone can be shown to generate spontaneous symmetry breaking and motions \cite{recho2013contraction}. 
        Finally, like other cell motility models \cite{shao2010computational,ziebert2012model,barnhart2015balance}, we have assumed that membrane tension is spatially uniform, yet spatial variations in tension has been observed in cells in experiments \cite{Shi2018-wa}.
        An extension of our approach to incorporate some of these additional details as well internal active stresses has the potential to further enrich the model.

\normalem
\bibliographystyle{unsrt}
\bibliography{pnassample.bib}

\end{document}